\documentclass[aps,prd,showpacs,twocolumn,10pt,superscriptaddress,floatfix,nofootinbib]{revtex4-1}

\usepackage{amssymb,amsfonts,amsmath,bm,bbm}
\usepackage{graphicx}
\usepackage{dcolumn}
\usepackage{bm}
\usepackage[
colorlinks=true,
breaklinks=true,
citecolor=blue]{hyperref}
\usepackage{bbold}
\usepackage{epstopdf}

\newcommand{\itp}{\affiliation{CAS Key Laboratory of Theoretical Physics,
            Institute of Theoretical Physics,\\ Chinese Academy of Sciences,
            Beijing 100190, China}}

\newcommand{\lzu}{\affiliation{School of Nuclear Science and Technology, Lanzhou University, Lanzhou 730000, China}}

\begin{document}

\title{Effect of the chiral phase transition on axion mass and self-coupling}

\author{Zhen-Yan~Lu}%
 \email{luzhenyan@itp.ac.cn}\itp

\author{Marco Ruggieri}%
 \email{Corresponding author: ruggieri@lzu.edu.cn}\lzu


\date{\today}

\begin{abstract}
We compute the effect of the chiral phase transition of QCD
on the axion mass and self-coupling; the coupling of the axion to the quarks at finite temperature is described within the
Nambu$-$Jona-Lasinio model.
We find that the axion mass  decreases with temperature, following the response of the topological susceptibility,
in agreement with previous results obtained within chiral perturbation theory at low  and intermediate temperatures.
As expected, the comparison with lattice data shows that chiral perturbation theory fails to reproduce the topological susceptibility
around the chiral critical temperature, while the Nambu$-$Jona-Lasinio model offers a better qualitative agreement with these data,
hence a more reliable estimate of the temperature dependence of the axion mass in the presence of a hot quark medium.
We complete our study by computing the temperature dependence of the self-coupling of the axion,
finding that this coupling decreases at and above the phase transition.
The model used in our work as well as the results presented here
pave the way to the computation of the in-medium effects of hot and/or dense quark-gluon plasma
on the axion properties.


\end{abstract}



\maketitle

\section{Introduction}

The current theory used to describe the strong interactions is QCD,
which possesses the $\mathrm{U}(1)_A$ anomaly as well as the spontaneous breaking of chiral symmetry
as some of its main features.
Because of the nontrivial topological structure of the QCD vacuum induced by the instanton effects,
a total derivative term is expected in the QCD Lagrangian,
\begin{equation}
{\cal L}_\theta \propto\theta F\cdot\tilde{F},\label{eq:U1aa}
\end{equation}
where $F$ and $\tilde{F}$ denote the gluonic field strength tensor and its dual, respectively, and the real parameter $\theta$
is known as the $\theta$ angle.
If $\theta\neq 0$, then QCD is not $CP$ symmetric;
it is, however, well known that the value of $\theta$ is very small, $\theta\lesssim 10^{-11}$;
see, for example, constraints from electric dipole moments \cite{06Baker.Doyle.ea131801-131801PRL,09Griffith.Swallows.ea101601-101601PRL,15Parker.Dietrich.ea233002-233002PRL,16Graner.Chen.ea161601-161601PRL,17Yamanaka.Yamada.ea65503-65503PRC}
as well as lattice QCD calculations~\cite{15Guo.Horsley.ea62001-62001PRL,15Bhattacharya.Cirigliano.ea212002-212002PRL}.
The very small value of $\theta$ implies that strong interactions conserve $CP$ remarkably well.
To understand why QCD is $CP$ conserving despite the possibility of a $CP$-breaking term in its Lagrangian, a new global, chiral $\mathrm{U}(1)_{PQ}$ symmetry was added to the QCD Lagrangian, and then an additional $CP$-violating term originating from the $\mathrm{U}(1)_{PQ}$ anomaly would exactly eliminate the $\theta$ term above.
This mechanism was proposed by Peccei and Quinn~\cite{77Peccei.Quinn1440-1443PRL,77Peccei.Quinn1791-1797PRD}
and  is often referred to as the Peccei-Quinn (PQ) mechanism.
Soon after this proposal, it was pointed out \cite{78Weinberg223-226PRL,78Wilczek279-282PRL} that a  pseudo-Goldstone boson,
namely, the QCD axion (which we will call simply the axion in the following),
would arise from the spontaneous symmetry breaking of the $\mathrm{U}(1)_{PQ}$ symmetry;
see also Refs.~\cite{16Cortona.Hardy.ea34-34JHEP,10Kim.Carosi557-601RMP,11Mack21-21JCAP,15Berkowitz.Buchoff.ea34507-34507PRD}.

Axions are weakly interacting and  very light particles and thus are good cold dark matter
candidates~\cite{09Duffy.Bibber105008-105008NJP,78Weinberg223-226PRL,91Turner.Wilczek5-8PRL,09Visinelli.Gondolo35024-35024PRD}.
Moreover, it has been suggested that they can form stars~\cite{91Tkachev289-293PLB,93Kolb.Tkachev3051-3054PRL,11Chavanis43531-43531PRD,06Guzman.Urena-Lopez814-819AJ,11Barranco.Bernal43525-43525PRD,16Braaten.Mohapatra.ea121801-121801PRL,16Davidson.Schwetz123509-123509PRD,16Eby.Leembruggen.ea66-66JHEP,17Helfer.Marsh.ea55-55JCAP,17Levkov.Panin.ea11301-11301PRL,17Eby.Leembruggen.ea14-14J,18Visinelli.Baum.ea64-72PLB,16Chavanis83007-83007PRD,16Cotner63503-63503PRD,16Bai.Barger.ea127-127JHEP}
as well as a
Bose-Einstein condensate~\cite{09Sikivie.Yang111301-111301PRL}
with a very high condensation temperature \cite{18Chavanis23009-23009PRD}.
It is therefore clear that it is important to know how the axion properties, in particular, the mass and the self-coupling,
evolve with temperature. This is the main scope of the present study, in which we compute how these two quantities
are affected by temperature, focusing, in particular, for temperatures around the critical temperature of QCD.

The study of the response of the axion to a QCD thermal medium in proximity of the critical temperature,
$T\approx 150$ MeV,
calls for the use of effective field theories and  phenomenological models, due to the impossibility of using 
perturbative QCD in this  moderate energy regime.
One of these effective theories is the chiral perturbation theory ($\chi$PT),
which is a systematic expansion in powers of the momenta of light mesons (namely, the pions for the
case of two-flavor QCD) and of the current quark masses;
$\chi$PT provides a good tool to study the $\theta$ vacuum of QCD;
see, for example, Refs.~~\cite{03Brower.Chandrasekharan.ea64-74PLB,09Mao.Chiu34502-34502PRD,10Aoki.Fukaya34022-34022PRD,12Bernard.Descotes-Genon.ea80-80JHEP,12Bernard.Descotes-Genon.ea51-51JHEP,06Metlitski.Zhitnitsky721-728PLB} as well as the QCD axion physics
at low temperature
\cite{16Cortona.Hardy.ea34-34JHEP}. In particular,
$\chi$PT predicts a value for the topological susceptibility at zero temperature \cite{16Cortona.Hardy.ea34-34JHEP}
that agrees with the lattice QCD results
~\cite{16Borsanyi.Fodor.ea69-71N,18Aoki4008-4008EWC,16Bonati.DElia.ea155-155JHEP}.
At finite temperatures, however, and, in particular, in the proximity of the QCD crossover,
the $\chi$PT results may become unreliable; from the physical point of view, this can be understood because
$\chi$PT is formulated in terms of pions and contains no information about the chiral crossover at high temperature,
around and above which a formulation in terms of quarks might be more appropriate.

Because of the limitations of $\chi$PT, in this work, we adopt the Nambu$-$Jona-Lasinio (NJL) model
to investigate the response of the axion to a finite temperature.
With the instanton effects taken into account, the NJL model provides a theoretical framework to simultaneously incorporate the spontaneous and explicit chiral symmetry breaking as well as the $\mathrm{U}(1)_A$ anomaly.
In fact, the interaction of the axion with quarks can be obtained by first adding the interaction term
\begin{equation}
{\cal L}_a = \theta F\cdot\tilde{F} + \frac{a}{f_a} F\cdot\tilde{F},\label{eq:U1aab}
\end{equation}
to the NJL Lagrangian, in which, according to the PQ mechanism, $\langle a/f_a+\theta \rangle= 0$;
then introducing the quantum fluctuation $a = \langle a \rangle + \delta a$
in the above equation, renaming $\delta a \rightarrow a$ where $a$
from now on denotes the axion field, and performing a chiral rotation that transfers
the interaction of $a$ with $F\cdot\tilde{F}$ to the interaction of the $a$ with the quarks.
After this chiral rotation is performed we are left with an effective theory of the axion interacting
with a thermal bath of quarks, the latter being capable of describing the important chiral crossover
of QCD at finite temperature, which instead lacks in $\chi$PT. In Eq.~(\ref{eq:U1aab}), $f_a$ is the axion decay constant, $10^8\lesssim f_a \lesssim 10^{12}$ GeV~\cite{18Takahashi.Yin.ea15042-15042PRD}.

The plan of the article is as follows.
In Sec.~\ref{NJL22} we present the NJL model augmented with the interaction with the instanton.
In Sec.~\ref{CHPT33} we briefy introduce the axion at finite temperature within the formalism of $\chi$PT.
In Sec.~\ref{RESULT} we report our results on the axion mass and self-coupling.
Finally, we present our conclusions and an outlook in Sec.~\ref{CONCLUSIONS}.

\section{Axion within the NJL model}\label{NJL22}

In this section we describe the model that we use in our calculations,
namely the NJL model with the 't Hooft term augmented by the interaction with the axion.
We use a two-flavor model in this article, while an extension to the three-flavor case will be the subject of a future study.
The NJL model Lagrangian incorporating the $\mathrm{U}(1)_A$ symmetry-breaking term
is given by
\begin{eqnarray}
\mathcal{L}=\bar{q}(i\gamma^\mu\partial_\mu-m)q
+\mathcal{L}_{\bar{q}q}+\mathcal{L}_\mathrm{det},
\end{eqnarray}
where $q$ denote the quark fields, $m$ is the current quark mass,
 and the attractive part of the $\bar{q}q$ channel of the Fierz
transformed color current-current interaction is given by
\begin{eqnarray}
\mathcal{L}_{\bar{q}q}=G_1[(\bar{q}\tau_aq)^2+(\bar{q}\tau_a i\gamma_5q)^2],
\end{eqnarray}
where $\tau_0=\mathcal{I}_{2\times 2}$ is the unit matrix and $\tau_i$ ($i=1,2,3$) denote the Pauli matrices.
Finally, we have put
\begin{eqnarray}
\mathcal{L}_\mathrm{det}=8G_2\left[e^{i\frac{a}{f_a}}\mathrm{det}(q_Rq_L)+e^{-i\frac{a}{f_a}}\mathrm{det}(q_Lq_R)\right],
\end{eqnarray}
which can be obtained by a chiral rotation of the quark fields in the path
integral \cite{76Hooft3432-3450PRD,86Hooft357-387PR,05Buballa205-376PR} starting from the
Lagrangian in Eq.(\ref{eq:U1aab});
in the above equation, the determinant is understood in the flavor space.
The determinant term breaks the original global symmetry,
$\mathrm{U}(2)_L\otimes \mathrm{U}(2)_R$, down to $\mathrm{SU}(2)_L\otimes \mathrm{SU}(2)_R\otimes \mathrm{U}(1)_B$.
The coupling constants $G_1$ and $G_2$ are often assumed to be equal in the literature.
This version of the NJL model~\cite{03Frank.Buballa.ea221-226PLB,09Boomsma.Boer34019-34019PRD,15Chatterjee.Mishra.ea34031-34031PRD,10Boomsma.Boer74005-74005PRD,01Fukushima.Ohnishi.ea45203-45203PRC,13Xia.He.ea56013-56013PRD}, as well as that enhanced by the Polyakov loop
\cite{11Sakai.Kouno.ea349-355PLB} have been widely used to investigate the $\theta$ effects on the QCD phase transition.

To obtain the thermodynamic potential in the one-loop (often called the mean field) approximation,
we neglect the quantum fluctuation and replace the scalar and pseudoscalar fields with their corresponding condensates.
The final result is well known in the literature; therefore, we merely quote the result here, that is,
\begin{eqnarray}\label{Omega-all}
\Omega(\alpha_0,\beta_0)&=&\Omega_q-G_2(\eta^2-\sigma^2)\cos\frac{a}{f_a} \nonumber\\
&&+G_1(\eta^2+\sigma^2)-2G_2\sigma \eta \sin\frac{a}{f_a},
\end{eqnarray}
where $\sigma=\langle\bar{q}q\rangle$ and $\eta=\langle\bar{q} i\gamma_5q\rangle$ are the chiral and SU(2) isospin-singlet pseudoscalar condensates respectively. In the above equation the quark contribution reads
\begin{eqnarray}\label{OMEGA_q}
\Omega_q=-8N_c\int\frac{\mathrm{d}^3p}{(2\pi)^3}\bigg[\frac{E_p}{2}+T
\log\big(1+e^{- E_p/T}\big) \bigg],
\end{eqnarray}
where $N_c=3$ represents the number of color of quarks and
\begin{eqnarray}
E_p=\sqrt{p^2+M^2},~~M=\sqrt{(m+\alpha_0)^2+\beta_0^2}
\end{eqnarray}
is the single particle energy of quarks.
We have also introduced the condensates $\alpha_0$ and $\beta_0$, which are defined in terms of the standard QCD
ones as
\begin{eqnarray}
\alpha_0&=&-2\Big(G_1+G_2\cos\frac{a}{f_a}\Big)\sigma
+2G_2 \eta \sin\frac{a}{f_a},\\
\beta_0&=&-2\Big(G_1-G_2\cos\frac{a}{f_a}\Big) \eta
+2G_2\sigma \sin\frac{a}{f_a}.
\end{eqnarray}
In the following we put $G_1 = (1-c)G_0$ and $G_2 = c G_0$.

The integral in Eq.~(\ref{OMEGA_q}) can be split into two parts: a zero-temperature and a finite-temperature part, respectively,
corresponding to
the first and second terms in the right-hand side of Eq.~(\ref{OMEGA_q}). The zero-temperature contribution measures the energy difference
between the vacuum without condensation and the vacuum with condensation. This contribution is divergent
in the ultraviolet: in order to
handle this divergence, we follow the standard procedure and cut the integral at the scale $\Lambda$; on the other
hand, the thermal part is finite, and we do not regularize it.
Different regularization schemes of the zero-temperature part might lead to slightly different quantitative results,
but the qualitative picture is usually unchanged by changing this scheme; for this reason, we use only this simple
regularization here, leaving the study of different regularization schemes to future studies.

For a given value of $a$, the thermodynamic potential is a function of
$\sigma$ and $\eta$ or equivalently on $\alpha_0$ and $\beta_0$;
at each temperature, the physical values of the condensates
$\bar\sigma$ and $\bar\eta$
correspond to the solutions of the gap equations,
namely,
\begin{eqnarray}\label{Solution}
\frac{\partial \Omega}{\partial\sigma}\Big|_{\sigma=\bar{\sigma}}=0,~~
\frac{\partial \Omega}{\partial\eta}\Big|_{\eta=\bar{\eta}}=0.
\end{eqnarray}
At the physical point we can define the effective potential for the axion as
\begin{eqnarray}\label{VaPotential}
\mathcal{V}(a)=\Omega(\sigma=\bar{\sigma},\eta=\bar{\eta}|a).
\end{eqnarray}

The axion mass is defined in terms of
the second derivative of the effective potential at $a=0$~\cite{16Cortona.Hardy.ea34-34JHEP,15Kitano.Yamada136-136JHEP,16Borsanyi.Dierigl.ea175-181PLB,16Bonati.DElia.ea155-155JHEP,17Dine.Draper.ea95001-95001PRD}, that is,
\begin{eqnarray}\label{maNJLdefi}
m_a^2=\frac{\mathrm{d}^{2}\mathcal{V}(a)}{\mathrm{d}a^{2}}\Big|_{a=0}=\frac{\chi_t}{f_a^2},
\end{eqnarray}
where $\chi_t$ corresponds to the topological susceptibility;
similarly, the axion self-coupling is defined in terms of the fourth derivative of the effective potential at $a=0$, namely,
\begin{eqnarray}\label{Lambda}
\lambda_a=\frac{\mathrm{d}^{4}\mathcal{V}(a)}{\mathrm{d}a^{4}}\Big|_{a=0}.
\end{eqnarray}
The topological susceptibility plays an important role in understanding the physics of QCD vacuum as well as of the $\mathrm{U}(1)_A$ anomaly, which has been studied previously in the NJL model at finite temperature.
It is worth remarking that
the derivative in Eq.~(\ref{Lambda}) is understood as a total derivative that takes into
account the fact that the condensates may have a dependence on $a$, namely,
\begin{eqnarray}
\frac{\mathrm{d}\mathcal{V}}{\mathrm{d}a}
=\frac{\partial\mathcal{V}}{\partial a}+\frac{\partial\mathcal{V}}{\partial\sigma}
\frac{\partial\sigma}{\partial a}+\frac{\partial\mathcal{V}}{\partial\eta}
\frac{\partial\eta}{\partial a}.
\end{eqnarray}

\section{Axion within $\chi$PT}\label{CHPT33}
We briefly review here the axion at finite temperature within $\chi$PT; the results shown here are well known so we
refer to original literature for a detailed discussion (see, for example, Refs.~
\cite{87Gasser.Leutwyler83-88PLB,89Gerber.Leutwyler387-429NPB,16Cortona.Hardy.ea34-34JHEP}).
In the SU(2) $\chi$PT framework, the temperature dependence of the axion potential up to the next-to-leading order (NLO) is
given by
\begin{eqnarray}
\mathcal{V}(a,T)&=&\mathcal{V}_0(a)\bigg[1-\frac{3}{2}\frac{T^4}{\pi^2f_\pi^2M_a^2} \nonumber\\
&&\times
\int_0^\infty
 x^2\log\big(1-e^{-E_a}\big)\mathrm{d}x\bigg],~~\label{eq:pot1}
\end{eqnarray}
where $f_\pi$ is the pion decay constant and $\mathcal{V}_0$ corresponds to
the NLO axion potential at zero temperature \cite{16Cortona.Hardy.ea34-34JHEP};
moreover, $E_a=\sqrt{x^2+M_a^2/T^2}$ with $M_a$ being the leading-order pion mass in a nonvanishing axion background,
which in the $m_u = m_d$ limit reads \cite{03Brower.Chandrasekharan.ea64-74PLB,15Guo.Meisner278-282PLB,16Cortona.Hardy.ea34-34JHEP}
\begin{eqnarray}
M_a^2=m_\pi^2\cos\frac{a}{2f_a}.
\end{eqnarray}
The temperature-dependent axion mass can be obtained easily by taking the second derivative of the potential
in Eq.~(\ref{eq:pot1}), that is
\begin{eqnarray}\label{maRatio}
\frac{m_a^2(T)}{m_a^2}&=&1-\frac{3T^2}{4\pi^2f_\pi^2}\int_0^\infty \frac{x^2}{E_T} \frac{\mathrm{d}x}{e^{E_T}-1} ,~~~
\end{eqnarray}
where $E_T=\sqrt{x^2+m_\pi^2/T^2}$ and we have put $m_a = m_a(T=0)$.
Similarly the self-coupling at finite temperature is easily obtained,
\begin{eqnarray}\label{LaRatio}
\frac{\lambda_a(T)}{\lambda_a}&=&1-\frac{3T^2}{4\pi^2f_\pi^2}\int_0^\infty \frac{x^2}{E_T} \frac{\mathrm{d}x }{e^{E_T}-1}
 \nonumber\\
&&+\frac{9m_\pi^2}{8f_\pi^2}\int_0^\infty \frac{x^2}{\pi^2}\frac{e^{E_T}(E_T+1)-1}{( e^{E_T}-1)^2E_T^3}\mathrm{d}x, ~~~~~~
\end{eqnarray}
where $\lambda_a = \lambda_a(T=0)$.
The pion mass and the decay constant are experimentally well known~\cite{16Patrignani.others100001-100001CPC,14Bijnens.Ecker149-174ARNPS}; therefore,
the uncertainties on the axion mass and self-coupling at zero temperature come from the quark mass ratio and the renormalized NLO couplings~\cite{16Cortona.Hardy.ea34-34JHEP}.
From Eqs.~(\ref{maRatio}) and (\ref{LaRatio}), we notice, however, that the ratio $m_a^2(T)/m_a^2$ and $\lambda_a(T)/\lambda_a$ are independent of the
quark masses and the NLO couplings. This implies that the ratios can be evaluated with better precision
than the axion mass and self-coupling at zero temperature within $\chi$PT, at least when the temperature is much lower than the QCD critical temperature.

\section{Results}\label{RESULT}

In this section we summarize the results for the axion mass and its self-coupling obtained within the NJL model around and above the
QCD critical temperature, and we compare these with the same quantities computed within $\chi$PT.
The parameters of the NJL model  are those of Refs.~\cite{03Frank.Buballa.ea221-226PLB,09Boomsma.Boer34019-34019PRD},
that is, $\Lambda$ = 590 MeV,
$G_0$ = $2.435/\Lambda^2$, $c=0.2$ and $m$ = 6 MeV:
they are fixed by fitting the physical pion mass $m_\pi=140.2$ MeV,
the pion decay constant $f_\pi=92.6$ MeV, and the chiral condensate at zero temperature
$\sigma_0=2(-241.5~\mathrm{MeV})^3$.
When we compare the NJL results with the $\chi$PT ones, for the latter we use the parameters of Ref.~\cite{16Cortona.Hardy.ea34-34JHEP}.
For completeness, we will first present shortly some result about the axion potential as well as the topological susceptibility at finite temperature.

\subsection{Effective potential and topological susceptibility}

\begin{figure}
  \includegraphics[width=246pt]{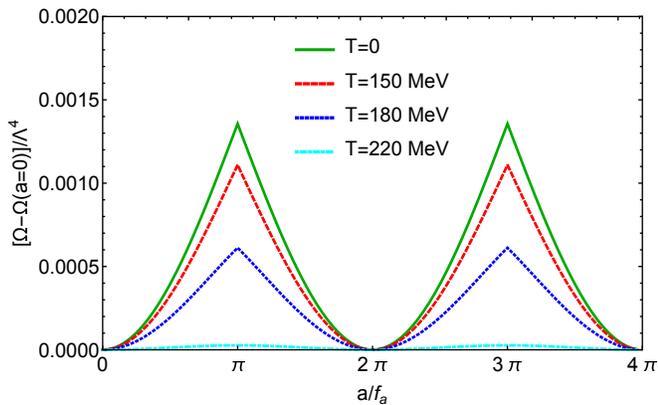}\\
  \caption{Effective potential for several values of the temperature. At each temperature, the potential is measured with
respect to the potential at $a=0$.}\label{OmegaTheta}
\end{figure}

\begin{figure}
  \includegraphics[width=246pt]{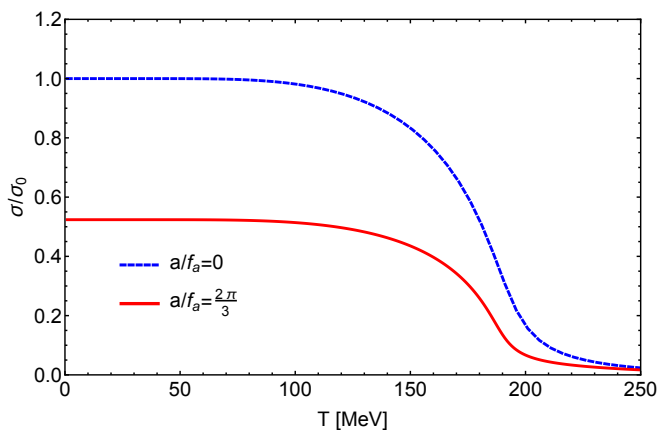}\\
  \caption{The chiral condensate $\sigma$, scaled by the corresponding zero-temperature value $\sigma_0$  in the vacuum, as a function of the temperature for $a/f_a=0$ and $2\pi/3$.}\label{sigmaSIGMA0}
\end{figure}

In Fig.~\ref{OmegaTheta} we plot the effective potential defined in Eq.~(\ref{VaPotential})
as a function of $a/f_a$, for several values of the temperature, computed within the NJL model.
At each temperature, we have subtracted the value of the potential at $a$ = 0.
We notice that at zero temperature the
effective potential shows a valley-hill structure, with degenerate vacua at $a/f_a$ = 0mod($2\pi$) and local maxima
at $a/f_a$ = $\pi$mod($2\pi$). This potential attains a minimum at $a=0$ in agreement with the Vafa-Witten theorem.
On the other hand, as the temperature is increased up to, and above, the critical temperature for
chiral symmetry restoration,
the effective potential becomes flatter, reflecting the suppression of the
height of the potential barrier at finite temperature.
The effect of the temperature described here is in qualitative agreement
with that of the magnetic field~\cite{15Chatterjee.Mishra.ea34031-34031PRD}.

In Fig.~\ref{sigmaSIGMA0} we show the ratio of the chiral condensate to the value in the vacuum for two typical values of $a$. Clearly the chiral condensate depends both on temperature and $a$ now. In the considered temperature range the chiral condensates all decrease monotonously with increasing temperature, reflecting the effective restoration of the chiral symmetry.

\begin{figure}
  \includegraphics[width=246pt]{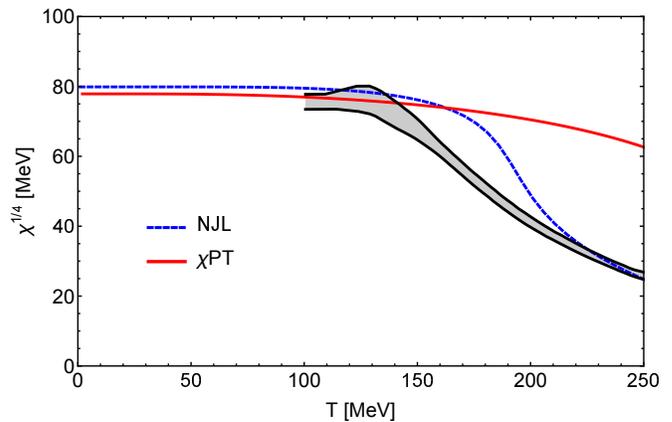}\\
  \caption{Fourth root of the topological susceptibility obtained from several methods as a function of the temperature. The shadow area represents the lattice data  taken from Ref.~\cite{16Borsanyi.Fodor.ea69-71N}.}\label{chiT}
\end{figure}

Next, we turn to the topological susceptibility.
First, we notice that from Eq.~(\ref{maNJLdefi}) the topological susceptibility at zero temperature can be obtained
within the NJL model as
$
\chi_t^{1/4}=79.87~\mathrm{MeV}
$
which is in fair agreement with Ref.~\cite{12Gatto.Ruggieri54013-54013PRD}
as well as with $\chi$PT~\cite{16Cortona.Hardy.ea34-34JHEP}
$
\chi_t^{1/4}=77.8(4)~\mathrm{MeV}
$
and lattice simulations~\cite{16Borsanyi.Fodor.ea69-71N}
$
\chi_t^{1/4}=78.1(2)~\mathrm{MeV}
$ in the isospin symmetric case.
In Fig.~\ref{chiT}, we plot (the fourth root of) the topological susceptibility
as a function of temperature; we show the result obtained within the NJL model (dashed blue line),
$\chi$PT (solid red line), and lattice simulations (shadow area).

We notice that both the NJL model and $\chi$PT are in agreement with lattice data up to
$T\lesssim 140$ MeV,
namely up to approximately the pseudocritical temperature of QCD.
Qualitatively, a difference between NJL and $\chi$PT is observed at higher temperatures:
the NJL contains the information of the partial chiral symmetry restoration at finite temperature;
therefore, it is capable of reproducing at least qualitatively the behavior of the topological susceptibility
measured in lattice QCD. On the other hand, $\chi$PT contains no information about chiral symmetry restoration
at large temperature, and this leads to a big discrepancy of this effective theory with QCD when the temperature
is approximately equal to or larger than the critical temperature.

\subsection{Axion mass and its self-coupling constant}

\begin{figure}
  \includegraphics[width=246pt]{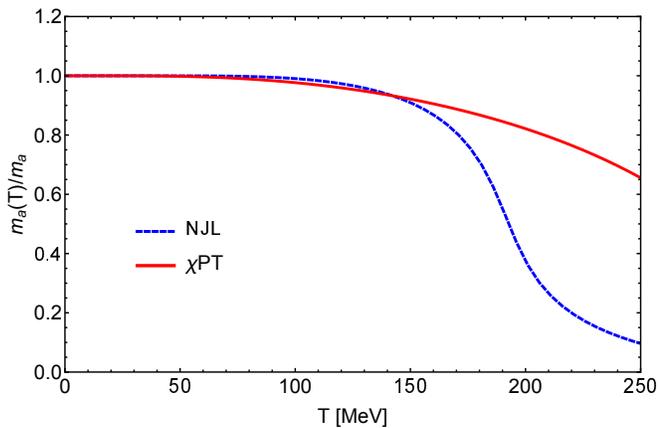}\\
  \caption{The thermal behavior of the temperature dependence of the axion mass from the NJL model scaled by its zero-temperature value. For comparison, we also show the results from the $\chi$PT~\cite{16Cortona.Hardy.ea34-34JHEP} and recent lattice data~\cite{16Borsanyi.Fodor.ea69-71N}.}\label{maTmaT0}
\end{figure}

We now discuss the axion mass and its self-coupling within the NJL model at finite temperature.
It is worth reminding the reader here that the mass is an interesting quantity, for example,
in cavity microwave experiments that aim to detect  axions by stimulating their conversion to photons in a strong magnetic field~\cite{03Bradley.Clarke.ea777-817RMP,10Kim.Carosi557-601RMP,16Marsh1-79PR}.
From Eq.~(\ref{maNJLdefi}),
the axion mass at zero temperature within the NJL model is
\begin{eqnarray}
m_a=6.38\times\frac{10^3}{f_a}~\mathrm{MeV}^2,
\end{eqnarray}
in agreement with the result of $\chi$PT, $m_a=6.06(5)\times10^3/f_a~\mathrm{MeV}^2$ in the isospin symmetric
case,
as well as with that of the invisible axion model, $m_a\simeq6.0\times10^3/f_a~\mathrm{MeV}^2$~\cite{87Kim1-177PR,88Cheng1-89PR,90Turner67-97PR,90Raffelt1-113PR}.
The axion self-coupling
might play some role in the formation of the so-called axion stars~\cite{16Braaten.Mohapatra.ea121801-121801PRL,16Bai.Barger.ea127-127J}.
At zero temperature, this  can be computed within the NJL model, namely,
\begin{eqnarray}
\lambda_a=-\left(\frac{55.64~\mathrm{MeV}}{f_a}\right)^4,
\end{eqnarray}
in agreement with the $\chi$PT prediction, $\lambda_a=-(55.79(92)~\mathrm{MeV}/f_a)^4$~\cite{16Cortona.Hardy.ea34-34JHEP} in the case of two degenerate quark flavors.

In Fig.~\ref{maTmaT0}, we show the axion mass obtained within the NJL model (dashed blue line) and $\chi$PT (solid red line), scaled by their corresponding zero-temperature values, as a function of the temperature.
We find a rapid drop of $m_a$ around the QCD chiral crossover, $140~\mathrm{MeV}\lesssim T \lesssim 200~\mathrm{MeV}$,
within the NJL model,
which is just a different way to represent the rapid decrease of the topological susceptibility
in this temperature range.
On the contrary, the prediction of $\chi$PT for the axion mass is that this quantity is almost insensitive
to the chiral crossover and stays almost constant in the aforementioned temperature range,
a result in agreement with the behavior of the topological susceptibility discussed in the previous subsection.

\begin{figure}
  \includegraphics[width=246pt]{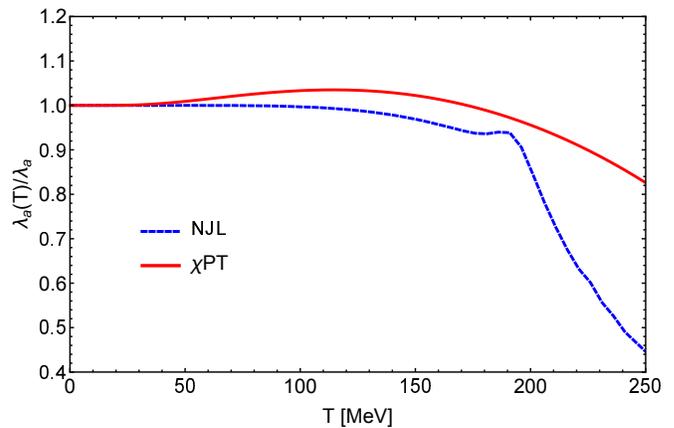}\\
  \caption{The thermal behavior of the axion self-coupling scaled by its zero-temperature value.}\label{LaTLa0Ratio}
\end{figure}

In Fig.~\ref{LaTLa0Ratio}, we plot the axion self-coupling normalized to its zero-temperature value,
as a function of the temperature; the lines and colors conventions are the same used in Fig.~\ref{maTmaT0}.
Despite the fact that $\chi$PT does not contain information about the chiral crossover,
we notice that the NJL model results
agrees with $\chi$PT well up to $T\approx 200$ MeV;
this suggests that the chiral crossover does not considerably affect the axion self-coupling.
Above this temperature,
the self-coupling from the NJL model experiences a quick drop, while that computed within $\chi$PT
experiences only a moderate decrease.

\section{Conclusions} \label{CONCLUSIONS}

In this article, we have studied the effect of the coupling of the QCD axion to a QCD-like thermal medium,
studying, in particular, the effect of the temperature and of the QCD crossover on its mass and its self-coupling.
The QCD medium at finite temperature has been described by the NJL model with two flavors,
which is capable of qualitatively reproducing  the chiral crossover of QCD;
we have compared the results obtained within the NJL model with those obtained previously
by means of $\chi$PT. The latter contains no information about the chiral crossover and is thus expected
to work well only for temperatures considerably smaller than the critical temperature of QCD;
this motivates the need to use another model, namely, the NJL model, to take into account the
(partial) restoration of chiral symmetry at finite temperature and compute the effect of this on
some phenomenological properties of the QCD axion.

We have found that below the critical temperature both the axion mass and the self-coupling do not show a substantial
temperature dependence. On the other hand,
the axion mass is very sensitive to the chiral crossover, showing a drop of its value within the crossover temperature range,
$140~\mathrm{MeV}\lesssim T \lesssim 200~\mathrm{MeV}$, an aspect that cannot be captured
by $\chi$PT since this contains no information about the critical temperature and is expected to be valid only
for temperatures much smaller than the critical temperature.
The axion self-coupling shows a less pronounced temperature dependence within both the NJL model and $\chi$PT.
We have found some substantial decrease of the coupling only for temperatures of the order of, or larger than, approximately $250$ MeV.
This means that if a hypothetical axion star is as hot as a young neutron
star~\cite{12Negreiros.Ruffini.ea12-12AA},  $T\lesssim 10^{10} \backsim 10^{11}$ K,
then the temperature dependence of the self-coupling can be ignored;
however, for a medium with a much higher temperature~\cite{16Bai.Barger.ea127-127JHEP,18Chavanis23009-23009PRD},
this temperature dependence should be taken into account.

We want to conclude this article by remarking that our main objective has not been that of performing a complete
phenomenological analysis of the low-energy properties of the QCD axion; instead,
we have pointed out that it is possible to study the interaction of this elusive particle with a QCD hot medium
by using an effective QCD model that is capable of capturing at least the qualitative aspects of the QCD phase diagram,
in particular the existence of a smooth crossover to a high-temperature phase in which chiral symmetry is approximately
restored. This cannot be taken into account by using $\chi$PT since the latter does not contain any information about
the QCD structure at high temperature; therefore, the results obtained within $\chi$PT are reliable only for temperatures
lower than the QCD critical temperature, while the present work aims to extend the study of thermal properties
of the QCD axion up to, and beyond, the QCD critical temperature.
For a possible outlook of the research presented here, we mention that, keeping in mind the well-known limitations of the NJL model,
it is possible to extend this study to other contexts: for example,
it is interesting to compute in-medium properties of the axion coupled to
hot and dense quark matter as well as to derive the temperature and density dependence of the low-energy
axion Lagrangian. We plan to report on these topics in the future.

\section{Acknowledgments}

We thank Jing-Yi Chao, Feng-Kun Guo, 
Sergi Gonz\`alez-Sol\'is, and John Petrucci for valuable discussions and inspiration.
This work is supported in part by DFG and NSFC through funds provided
to the Sino-German CRC 110 ``Symmetries and the Emergence of Structure in QCD'' (NSFC
Grant No.~11621131001 and DFG Grant No.~TRR110), by NSFC (Grant
No.~11747601),
by the Young 1000 Talents Program of China,
 by the CAS Key Research Program of Frontier
Sciences (Grant No.~QYZDB-SSW-SYS013), and by the CAS President's International
Fellowship Initiative (PIFI) (Grant No.~2017VMA0025).
This work is also in part supported by China Postdoctoral Science Foundation
(Grant No.~2017M620920).  The work of M. R.  is supported by the National Science Foundation of China
(Grant No.~11875153) and by the Fundamental Research Funds for the
Central Universities (Grant No.~862946).

\bibliographystyle{aapmrev4-2}  
\bibliography{MyRefBB}


\end{document}